\documentclass[apj,square,numbers]{emulateapj}

\usepackage{graphicx}
\usepackage{xcolor}
\usepackage{amsmath,amssymb,latexsym}
\usepackage{hyperref}
\usepackage{ulem}

\begin{document}
\def\gsim{ \lower .75ex \hbox{$\sim$} \llap{\raise .27ex \hbox{$>$}} }
\def\lsim{ \lower .75ex \hbox{$\sim$} \llap{\raise .27ex \hbox{$<$}} }

\title{Cosmology-independent estimate of the fraction of baryon mass in the IGM from fast radio burst observations}

\author{Zhengxiang Li$^{1}$, He Gao$^{1}$, Jun-Jie Wei$^{2}$, Yuan-Pei Yang$^{3,4}$, Bing Zhang$^{5}$, and Zong-Hong Zhu$^{1,6}$}

\affil{$^{1}$Department of Astronomy, Beijing Normal University, Beijing 100875, China; gaohe@bnu.edu.cn; zhuzh@bnu.edu.cn}
\affil{$^{2}$Purple Mountain Observatory, Chinese Academy of Sciences, Nanjing 210034, China;}
\affil{$^{3}$Kavli Institute for Astronomy and Astrophysics, Peking University, Beijing 100871, China;}
\affil{$^{4}$National Astronomical Observatories of China, Chinese Academy of Sciences, Beijing 100012, China;}
\affil{$^{5}$Department of Physics and Astronomy, University of Nevada, Las Vegas, NV 89154, USA;}
\affil{$^{6}$School of Physics and Technology, Wuhan University, Wuhan 430072, China}

\begin{abstract}
The excessive dispersion measure (DM) of fast radio bursts (FRBs) has been proposed to be a powerful tool to study intergalactic medium (IGM) and to perform cosmography. One issue is that the fraction of baryons in the IGM, $f_{\rm IGM}$, is not properly constrained. Here we propose a method of estimating $f_{\rm IGM}$ using a putative sample of FRBs with the measurements of both DM and luminosity distance $d_{\rm L}$. The latter can be obtained if the FRB is associated with a distance indicator (e.g. a gamma-ray burst or a gravitational wave event), or the redshift $z$ of the FRB is measured and $d_{\rm L}$ at the corresponding $z$ is available from other distance indicators (e.g. type Ia supernovae) at the same redshift. Since $d_{\rm L}/{\rm DM}$ essentially does not depend on cosmological parameters, our method can determine $f_{\rm IGM}$ independent of cosmological parameters.  We parameterize $f_{\rm IGM}$ as a function of redshift and model the DM contribution from a host galaxy as a function of star formation rate. Assuming  $f_{\rm IGM}$ has a mild evolution with redshift with a functional form and by means of Monte Carlo simulations, we show that an unbiased and cosmology-independent estimate of the present value of $f_{\rm IGM}$ with a $\sim 12\%$ uncertainty can be obtained with 50 joint measurements of $d_{\rm L}$ and DM. In addition, such a method can also lead to a measurement of the mean value of DM contributed from the local host galaxy.

\end{abstract}
\maketitle
\renewcommand{\baselinestretch}{1.5}

\section{Introduction}
Fast radio bursts (FRBs) are a class of bright transients with millisecond durations detected at $\sim\mathrm{GHz}$ \citep{lorimer07,thornton13,petroff15,petroff16}. Although the progenitors of these events are debated, the large values of the dispersion measure (DM) in excess of the Galactic value of these events \citep{dolag15} suggest their extragalactic or even cosmological origin. The localization of the repeating burst FRB 121102 at $z \sim 0.19$ has firmly established the cosmological origin of at least this event \citep{spitler16,scholz16,chatterjee17,marcote17,tendulkar17}. The observed DM of an FRB therefore should at least have a significant contribution from the intergalactic medium (IGM). Based on the average DM-$z$ relation, one can estimate that the FRBs are typically at distances of the order of
gigaparsec \citep{ioka03,inoue04,deng14,zheng14,zhang18}. The cosmological origin of FRBs allows them to become promising tools to study the universe and fundamental physics, e.g. probing baryon density and spatial distribution \citep{mcquinn14,deng14}, constraining dark energy equation of state \citep{gao14, zhou14, walters18}, probing ionization history of the universe \citep{deng14,zheng14}, tracing the large-scale structure of the universe 
\citep{masui15}, testing the Einstein's equivalence principle \citep{wei15, nusser16, tingay16}, constraining the rest mass of the photon \citep{wu16,shao17}, measuring the cosmic proper distance \citep{wang17}, as well as constraining the magnetic fields in the IGM \citep{akahori16}. Thanks to another two properties of these mysterious transients, i.e. narrow durations ($\sim(1-10) \mathrm{ms}$) \citep{lorimer07, keane11, thornton13, spitler14, petroff15, ravi15, champion16, petroff16, spitler16} and high event rate ($\sim10^3-10^4$ per day all sky) \citep{thornton13, champion16}, lensed FRBs have been proposed as a probe of compact dark matter \citep{munoz16, wang18}, motion of the FRB source \citep{dai17}, as well as precision cosmology \citep{li18}.

In practice, there are some issues that hinder the application of FRBs for cosmological purposes. First, it is not easy to measure the redshifts of most FRBs. Localizing them is possible if they repeat, but it is possible that only a fraction of FRBs repeat \citep{palaniswamy18}. Recently, the Canadian Hydrogen Intensity Mapping Experiment (using the CHIME/FRB instrument) has reported detections of 13 FRBs during a pre-commissioning phase \citep{chime19a}. It is more exciting that one of the 13 FRBs is a second source of repeaters which suggests that CHIME/FRB and other wide-field sensitive radio telescopes will find a substantial population of repeating FRBs \citep{chime19b}. The second possibility is to hope that at least some FRBs are associated with other detectable catastrophic events, such as GRBs \citep{zhang14} and gravitational waves \citep{totani13,zhang16,wang16}, which is subject to future observational confirmations. The third hope would be to use large arrays to pin down the locations of FRBs, which may be possible with the operation of CHIME \citep{bandura14} and would be achievable with the operation of SKA \citep{macquart15}. 

Even if redshifts of FRBs are measured, there are two unknowns in connecting DM with the baryons in the IGM. One is the fraction of baryons in the IGM, i.e. $f_{\rm IGM}$. The distribution of baryons among stars, galaxies and IGM has been a subject of study. \citet{fukugita98} estimated that stars and their remnants comprise of about 17\% of the total baryons based on various observational constraints. Despite later intense studies based on numerical simulations \citep[e.g.][]{cen99,cen06} or observations \citep[e.g.][]{fukugita04,shull12,hill16,munoz18}, the IGM-fraction of baryons, $f_{\rm IGM}$ is still not well constrained. The second unknown is the local value of the DM from the FRB host galaxy, i.e. $\rm DM_{host,loc}$, which is related to the observed DM contribution from the host, $\mathrm{DM_{host}}$, via $\mathrm{DM_{host}}=\mathrm{DM_{host, loc}}/(1+z)$ \citep{ioka03,deng14}. The DM contributions from the galaxy itself have been studied \citep{xu15,luo18}. However, the free electrons near the FRB source may give significant contribution to DM. The degree of such a contribution is model dependent and very uncertain \citep{luan14,katz16,piro16,metzger17,piro17,cao17,yangzhang17}. Some methods of statistically constraining $\rm DM_{host,loc}$ have been proposed, and preliminary results suggest that the value could be relatively large at least for some FRBs \citep{yang16,yang17}.

\citet{gao14} and \citet{wei18} have noticed that making use the luminosity distance $d_{\rm L}$ (if available) of FRBs would be very helpful to perform cosmological tests. \citet{gao14} noticed that $d_{\rm L}/{\rm DM}$ is insensitive to dark energy equation of state, and \citet{wei18} noticed that $d_{\rm L} \cdot {\rm DM}$ is independent of the Hubble constant $H_0$. 

In this paper, assuming that a sample of FRBs can be localized so that their $d_{\rm L}$'s are measured, we propose that the measured $d_{\rm L}/{\rm DM}$ of the sample can be used as a powerful probe to achieve a nearly cosmology-free estimation for the fraction of baryons in the IGM. 
Through Monte Carlo simulations, we show that both the fraction of baryons in the IGM and the mean value of local host galaxy DM can be unbiasedly inferred from such joint measurements in a cosmological-model-independent way.

\section{Methods and Results}\label{sec2}

\subsection{Luminosity distance vs. dispersion measure}\label{sec2.1}
For an FRB with redshift measurement, there is a good chance to get its luminosity distance
\begin{equation}\label{eq1}
d_{\rm{L}}=\frac{c(1+z)}{H_0}\int_0^z\frac{dz'}{E(z')}.
\end{equation}
If the FRB is associated with a standardizable distance indicator, e.g. a GW or a GRB \citep[e.g.][]{totani13, zhang14, mingarelli15, wang16, zhang16, yamasaki17}, $d_{\rm L}$ can be directly measured. Even if FRBs are not associated with distance indicators, as long as its $z$ is measured, one may obtain its $d_{\rm L}$ using that of other indicators  (e.g. standard candles, rulers, and sirens) at the same redshift. In particular, there are a large number of well-measured type Ia supernovae (SNe Ia) at $z\lesssim2.3$.

The dispersion measure of an FRB is directly measured when it is discovered, which consists of the following three components \citep[e.g.][]{thornton13,deng14}
\begin{equation}\label{eq2}
\rm{DM}_{\rm{obs}}=\rm{DM}_{\rm{host}}+\rm{DM}_{\rm{IGM}}+\rm{DM}_{\rm{MW}},
\end{equation}
where subscripts ``host'', ``IGM'',  and ``MW'' denote contributions from the FRB host galaxy, IGM, and the Milky Way, respectively. The IGM portion of DM, i.e. $\rm{DM}_{\rm{IGM}}$ depends on the cosmological distance scale the burst passes through and the fraction of ionized electrons in hydrogen (H, $\chi_{\rm{e,H}}(z)$) and helium (He, $\chi_{\rm{e,He}}(z)$) on the path, which are closely related to the present-day baryon density parameter $\Omega_\mathrm{b}$ and fraction of baryons in the IGM \citep{deng14}. In reality, $f_{\rm{IGM}}$ is known to grow with redshift as massive haloes are less abundant in the early universe \citep{mcquinn14}. Here we parametrize the growth of $f_{\rm{IGM}}$ with redshift as a mildly increasing function, $f_{\rm{IGM}}(z)=f_{\rm{IGM,0}}(1+\alpha z/(1+z))$. The average value (for individual line of sight, the value may deviate from this due to the large scale density fluctuations \citep{mcquinn14}) can be written as
\begin{equation}\label{eq3}
\mathrm{DM}_{\mathrm{IGM}}=\frac{3cH_0\Omega_bf_{\mathrm{IGM,0}}}{8\pi G m_{\mathrm{p}}}\int_0^z\frac{\chi(z')(1+z'+\alpha z')dz'}{E(z')},
\end{equation} 
where
\begin{align}
\chi(z)&=\frac{3}{4}y_1\chi_\mathrm{{e,H}}(z)+\frac{1}{8}y_2\chi_\mathrm{{e,He}}(z), \nonumber\\ 
E(z)&=[\Omega_m(1+z)^3+(1-\Omega_m)^{3(1+w)}]^{1/2}. \nonumber
\end{align}
Here $\Omega_\mathrm{b}$ is the present-day baryon density parameter, $y_1\sim1$ and $y_2\simeq4-3y_1\sim1$ are the H and He mass fractions normalized to the typical values 3/4 and 1/4, respectively.

\begin{figure*}[htbp]
		\centering
	\includegraphics[width=0.45\textwidth, height=0.270\textwidth]{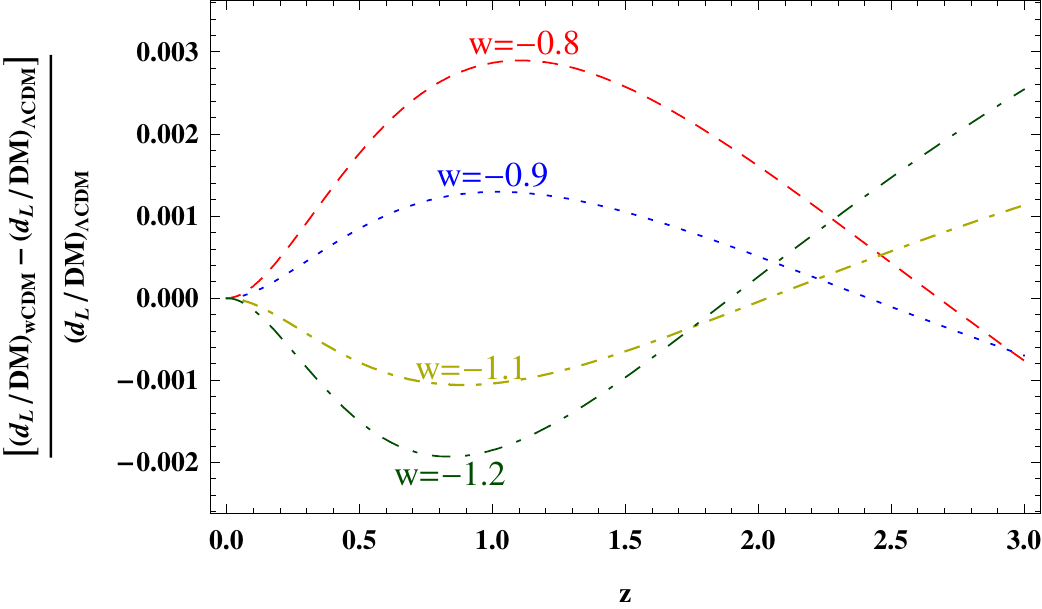}	
	\includegraphics[width=0.45\textwidth, height=0.270\textwidth]{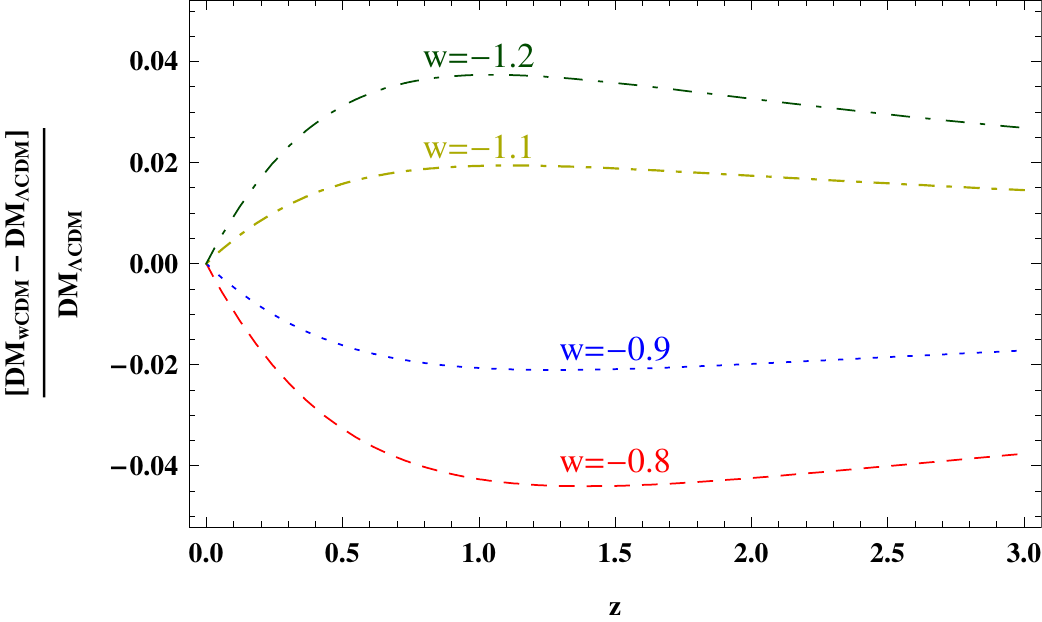}
	\caption{\label{Fig1} {\bf Left: }Deviations of the $d_{\mathrm{L}}$/DM inferred in the $w$CDM with different $w$ from the one inferred in the standard $\Lambda$CDM. {\bf Right: }Deviations of the ratio DM inferred in the $w$CDM with different $w$ from the one inferred in the standard $\Lambda$CDM.}
\end{figure*}

As suggested in \citet{gao14}, the ratio $d_{\mathrm{L}}$/DM is nearly insensitive to the dark energy equation of state $w$. That is, the variation of $d_{\mathrm{L}}$/DM resulting from the uncertainty of cosmological parameters is negligible. For the sake of comparison, in Figure \ref{Fig1}, we show deviations of both $d_{\mathrm{L}}$/DM and DM inferred in the $w$CDM with different $w$ from the ones inferred in the standard $\Lambda$CDM. It is suggested that, compared with the regular DM-$z$ relation, the sensitivity of $d_{\mathrm{L}}$/DM to cosmological parameters can be decreased by more than a factor of ten. As a result, this ratio can lead to an almost cosmology-free estimation for $f_{\rm {IGM}}$. In addition to statistical errors of cosmological parameters constrained from observations, the well-known Hubble constant tension, which is the $>3\sigma$ inconsistency between the expansion rate directly obtained from local distance measurements \citep{riess16} and the one constrained from high redshift cosmic microwave background radiation observations \citep{planck18}, implies the uncertainty of the standard $\Lambda$CDM itself. Therefore, any cosmological-model-independent methods or those with much less sensitivities to cosmological parameters for estimating $f_{\rm {IGM}}$ are worthy of a consideration.

\subsection{Redshift distribution and uncertainties of the parameters}

In order to generate a mock sample of FRBs to test our suggestion, one needs to assume the redshift distribution of FRBs and the uncertainties of the parameters. Without identifying the progenitor systems of FRBs, one cannot identify the redshift distribution. On the other hand, our study essentially does not depend on the true redshift distribution, since one cares about the DM at a certain redshift rather than how many bursts are there at a certain redshift. As a result, we adopt a simply phenomenological model by assuming a constant comoving number density of FRB, with the introduction of a Gaussian cutoff at some redshift $z_{\rm{cut}}$ to represent the decrease of the detected FRBs beyond it due to the instrumental signal-to-noise threshold effect \citep[e.g.][]{munoz16, li18}. Then, one has
\begin{equation}\label{eq4}
N_{\rm{const}}(z)=\mathcal{N}_{\rm{const}}\frac{\tilde{\chi}^2(z)}{H(z)(1+z)}e^{-{d_{\rm{L}}}^2(z)/[2{d_{\rm{L}}}^2(z_{\rm cut})]},
\end{equation}
where $\mathcal{N}_{\rm{const}}$ is a normalization factor to ensure that the integration of  $N_{\rm{const}}(z)$ is unity, $\tilde{\chi}(z)$ is the comoving distance, and $H(z)$ is the Hubble parameter at redshift $z$. In view of the detected FRBs \citep{petroff16} and their DM values, one may estimate their redshift distribution using the DM-$z$ relation \citep{deng14}. Even though some FRBs up to $z \sim 3$ may have been detected \citep{zhang18}, in general there is a decrease of FRB numbers beyond $z \sim 1$. So we adopt  $z_{\rm cut}=1$ in Equation \ref{eq4} for our further simulations.

\begin{figure*}[htbp]
	\centering
	\includegraphics[width=0.45\textwidth, height=0.25\textwidth]{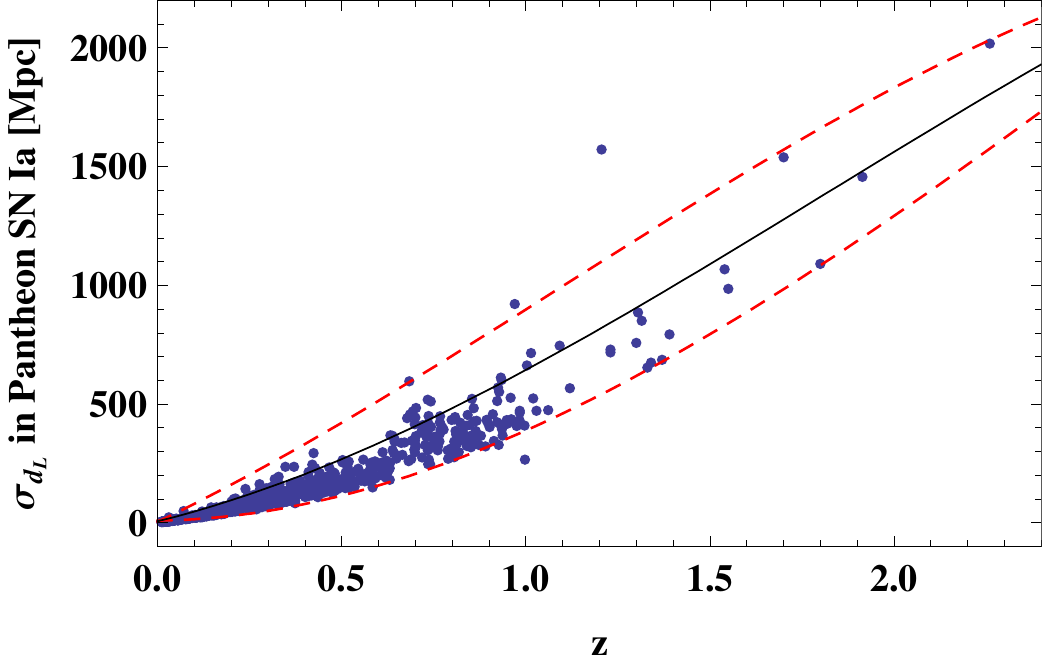}
	\includegraphics[width=0.45\textwidth, height=0.26\textwidth]{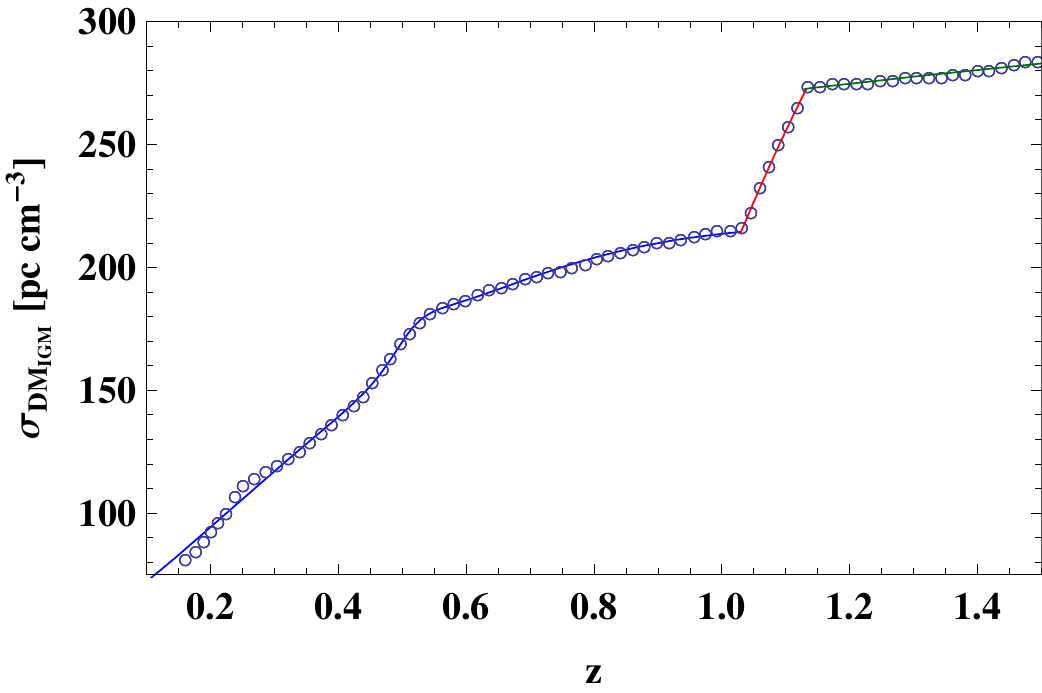}
	\caption{\label{Fig2} {\bf Left: }Uncertainty level of luminosity distances in the latest Pantheon SNe Ia sample. {\bf Right: } The DM uncertainty $\sigma_{\mathrm{IGM}}(z)$ derived from the simulations presented in \citet{mcquinn14}. }
\end{figure*}

The uncertainties of the latest Pantheon SN Ia \citep{scolnic18}, which consists of 1048 well-measured events, can stand for the typical level of precision in current luminosity distance measurements. The $\sigma_{d_{\rm{L}}}$ values in the Pantheon SN Ia data set are plotted in the left panel of Figure \ref{Fig2}. Generally speaking, it shows a trend that errors increase with redshift except two outliers at $z=0.998$ and $z=1.206$. Moreover, we find that the uncertainties $\sigma_{d_{\rm{L}}}(z)$ are in the region confined by two curves, $\sigma_{+}=3.687+744.097z+215.670z^2-65.343z^3$ and $\sigma_{-}=9.378+16.553z+412.942z^2-50.290z^3$ from above to below, respectively. Assuming that up-coming observations of $d_{\rm{L}}$ would also have uncertainties at this level, the curve $\sigma_0=(\sigma_{+}+\sigma_{-})/2$ can be used as an estimate of the $d_{\rm{L}}$ mean uncertainty. Then, the uncertainties of near-future $d_{\rm{L}}$ observations are random numbers $\tilde{\sigma}(z)$ drawn from the Gaussian distribution $N(\sigma_0(z), \epsilon(z))$, where the parameter $\epsilon(z)$ is chosen to be $(\sigma_{+}-\sigma_{-})/4$ so that $\tilde{\sigma}(z)$ falls in the confined area with a probability of 95.4\% \citep{ma11}.

For a well-localized FRB, we can subtract $\rm{DM_{MW}}$ from $\rm{DM_{obs}}$ with a considerable precision, and thus to get the extragalactic contribution \citep{yang16}
\begin{equation}\label{eq5}
\rm{DM_{ext}}=\rm{DM_{obs}}-\rm{DM_{MW}}=\rm{DM_{IGM}}+\rm{DM_{host}}.
\end{equation}
Here, we take $\rm{DM_{ext}}$ as the observed quantity, and its corresponding uncertainty is 
\begin{equation}\label{eq6}
\sigma_{\rm{ext}}=\big(\sigma_{\rm{obs}}^2+\sigma^2_{\rm{MW}}+\sigma^2_{\rm {IGM}}+\sigma_{\rm{host}}^2\big)^{1/2}.
\end{equation}
According to currently available observations compiled in the FRB catalog \citep{petroff16}, we obtain an average value $\sigma_{\rm {obs}}=1.5~{\rm {pc~cm^{-3}}}$ for the uncertainty of $\rm{DM_{obs}}$. For sources at high Galactic latitude ($|b| >10^{\circ}$), the average uncertainty of the DM contribution from the Milky Way is about $10~{\rm {pc~cm^{-3}}}$ \citep{manchester05}. The uncertainty of $\rm DM_{IGM}$ is more complicated because of the density fluctuation from the large scale structure. According to \citet{mcquinn14}, the standard deviation from the mean DM is dependent on the profile models characterizing the inhomogeneity of the baryon matter in the IGM. Here, we fit this numerical simulation results of  \citet{mcquinn14} and \citet{faucher-giguere11} and find that $\sigma_{\rm {IGM}}(z)$ can be described by a step function as shown in the right panel of Figure \ref{Fig2} (for both the simulation data and the fitting results). The value of $\rm {DM_{host}}$ and its uncertainty $\sigma_{\rm{host}}$ are intractable parameters since they are poorly known and dependent on many factors, such as the type of the host galaxy, the site of FRB in the host, the inclination angle of the galaxy disk, and the near-source plasma. In our analysis, we model the DM contribution from a host galaxy as a function with respect to redshift by assuming that the rest-frame DM distribution accommodates the evolution of star formation rate (SFR) history \citep{luo18}, 
\begin{equation}
{\rm DM_{host, loc}}(z)={\rm DM_{host, loc, 0}}\sqrt{\frac{{\rm SFR}(z)}{\rm {SFR}(0)}}.
\end{equation}
Here, for ${\rm SFR}(z)$, we adopt a continuous form of a broken power law \citep{hansan08},
\begin{equation}
{\rm SFR}(z)=0.02\bigg[(1+z)^{a\eta}+\bigg(\frac{1+z}{B}\bigg)^{b\eta}+\bigg(\frac{1+z}{C}\bigg)^{c\eta}\bigg]^{1/\eta},
\end{equation}
where $a=3.4,~b=-0.3$, and $c=-3.5$ are logarithmic slopes of the $0<z\leq1,~1<z\leq4$, and $z>4$ segments, respectively. The parameter $\eta\simeq-10$ is a smoothing factor for the transitions between different segments.

\subsection{Monte Carlo simulations and fitting results}\label{sec2.3}
With above preparations, we next investigate cosmological-model-independent constraints on the $f_{\rm {IGM}}$ by means of Monte Carlo simulations assuming that a sample of FRBs with joint measurements of $d_{\rm L}$ and DM is available. We adopt the standard spatially flat $\Lambda$CDM as the fiducial model with the parameters derived from the latest $Planck$ observations ($H_0=67.36\pm0.54~{\rm{km~s^{-1}~Mpc^{-1}}}$, $\Omega_m=0.315\pm0.007$, $\Omega_bh^2=0.02237\pm0.00015$) \citep{planck18}. In our simulations, we infer the fiducial values, $d^{\rm {fid}}_{\rm {L}}(z)$ and $\mathrm{DM}^{\rm {fid}}_{\mathrm{IGM}}(z)$, from Equations (\ref{eq1}) and (\ref{eq3}), respectively. The simulated luminosity distances are generated by $d^{\rm {sim}}_{\rm {L}}(z)=d^{\rm {fid}}_{\rm {L}}(z)+N(0,~\sigma_{d_{\rm{L}}}(z))$ with $\sigma_{d_{\rm{L}}}(z)\sim N(\sigma_0(z), \epsilon(z))$. For DM, we take $\rm DM_{ext}$ as the observed quantity. $\mathrm{DM}^{\rm {fid}}_{\mathrm{ext}}(z)$ is calculated from $\mathrm{DM}^{\rm {fid}}_{\mathrm{IGM}}(z)+{\rm {DM_{host, loc}}}(z)/(1+z)$ and ${\rm \langle{DM_{host, loc, 0}}}\rangle=N(150~{\rm pc~cm}^{-3}, 30~{\rm pc~cm}^{-3})$ is assumed to test the fidelity of the method. Then the simulated extragalactic dispersion measure is $\mathrm{DM}^{\rm {sim}}_{\mathrm{ext}}(z)=\mathrm{DM}^{\rm {fid}}_{\mathrm{ext}}(z)+N(0,~\sigma_{\rm {ext}}(z)$), where $\sigma_{\rm {ext}}(z)$ is obtained from Eq. (\ref{eq6}). In addition to cosmological parameters from the latest $Planck$ observations, we also take $f_{\rm {IGM,0}}=0.83$ \citep{fukugita98, shull12, deng14} and $\alpha=0.25$ as the fiducial values for DM simulations.

In order to obtain constraints of the fraction of baryon mass in IGM, $f_{\rm {IGM}}$, we need to separate $\rm DM_{IGM}$ from $\mathrm{DM}_{\mathrm{ext}}$, i.e. $\mathrm{DM}^{\rm {th}}_{\mathrm{ext}}(z; f_{\rm {IGM,0}}, \alpha, {\rm {DM_{host, loc,0}}})=\mathrm{DM}^{\rm{th}}_{\mathrm{IGM}}(z; f_{\rm {IGM,0}}, \alpha)+{\rm DM_{host, loc}}(z; {\rm {DM_{host, loc,0}}})/(1+z)$.
For a sample of FRBs with DM and $d_{\rm L}$ measurements, one can estimate the ability of cosmological-model-independent constraints on $f_{\rm {IGM}}$ from the combined parameter $d_{\rm{L}}/{\rm {DM}}$ by performing minimum $\chi^2$ statistics for the above-mentioned simulations. Moreover, estimations for the parameter $\alpha$ and the mean value of ${\rm {DM_{host, loc, 0}}}$, i.e. ${\rm \langle{DM_{host, loc, 0}}}\rangle$, can be obtained from the fitting in the meantime.

\begin{figure*}[htbp]
	\centering
	\includegraphics[width=0.315\textwidth, height=0.275\textwidth]{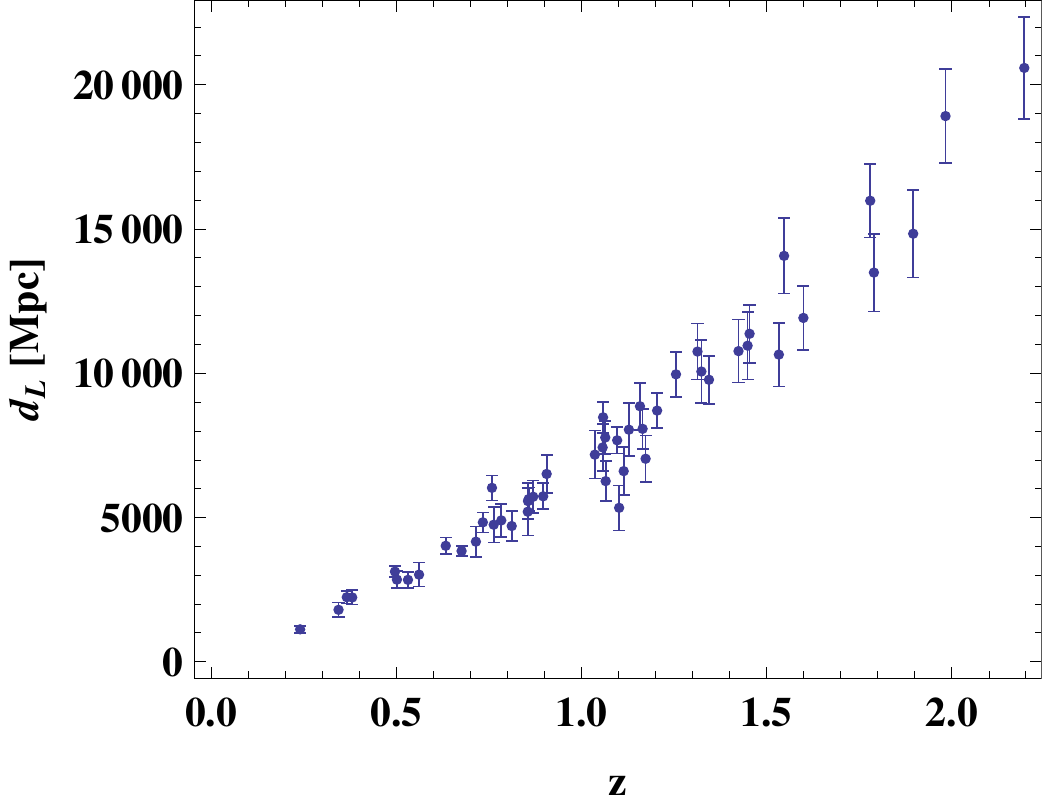}
	\includegraphics[width=0.315\textwidth, height=0.275\textwidth]{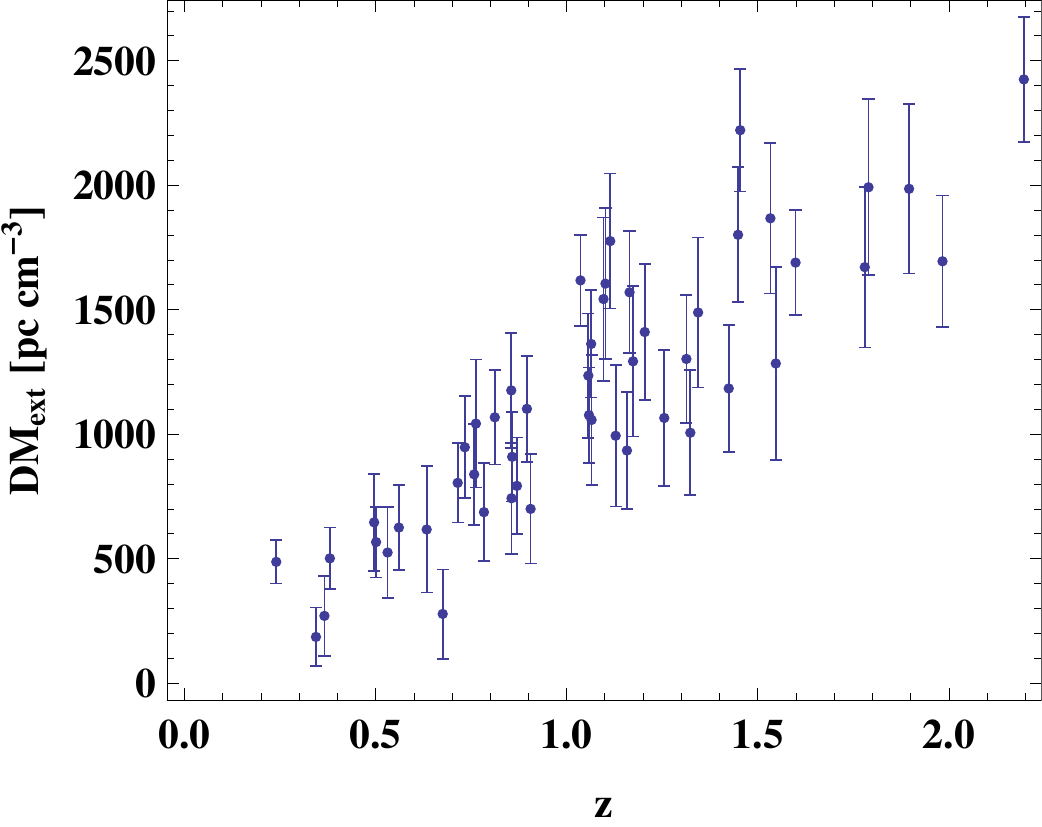}
	\includegraphics[width=0.33\textwidth, height=0.281\textwidth]{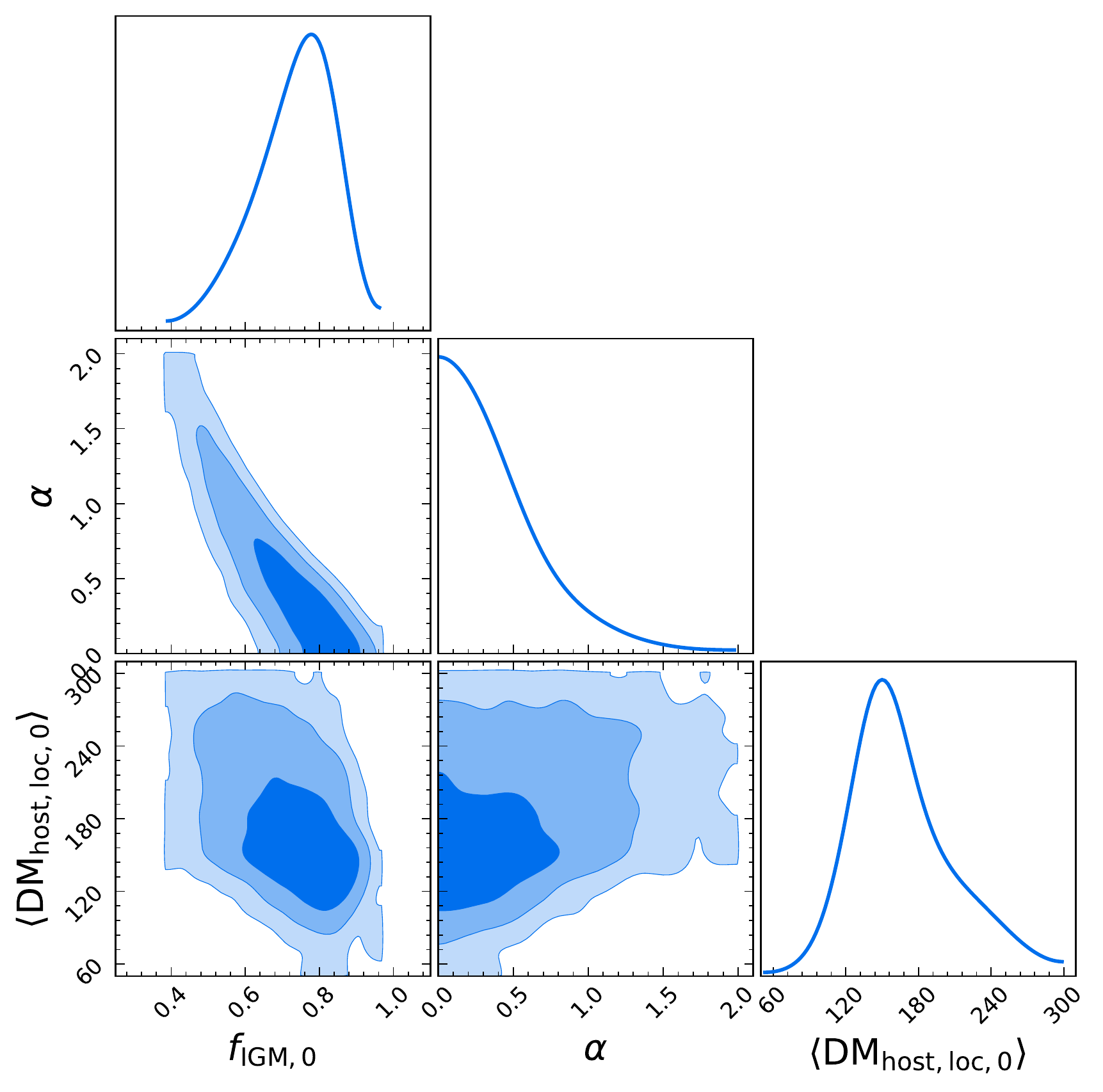}
	\caption{\label{Fig3} {\bf Left: } Mock data of 50 luminosity distances. {\bf Middle: } Mock data of 50 extragalactic dispersion measures. {\bf Right:} Constraints on $f_{\mathrm {IGM, 0}}$, $\alpha$, and $\langle\mathrm{DM}_{\mathrm{host, loc, 0}}\rangle$ from the corresponding simulations.}
\end{figure*}

We first consider the case with 50 joint measurements of $d_{\rm{L}}/{\rm {DM}}$ of FRBs. Simulated data sets for $d_{\rm{L}}$ and ${\rm {DM}}$ are shown in the left and middle panels of Figure \ref{Fig3}, respectively. We repeat these simulations for $\sim10000$ times and obtain the distributions of fitting results for $f_{\rm {IGM,0}}$, $\alpha$, and ${\rm \langle{DM_{host, loc, 0}}}\rangle$. They are shown in the right panel of Figure \ref{Fig3}. We obtain $f_{\rm {IGM}}=0.80^{+0.08}_{-0.12}$, $\alpha=0.01^{+0.61}_{-0.01}$, and ${\rm \langle{DM_{host, loc, 0}}}\rangle=153.78^{+56.10}_{-18.23}$. It suggests that an unbiased and cosmological-model-independent estimation for both the fraction of baryon mass in IGM and the mean value of DM host galaxy contribution can be expected as long as the sample size reaches 50. 

\begin{figure*}[htbp]
	\centering
	\includegraphics[width=0.315\textwidth, height=0.275\textwidth]{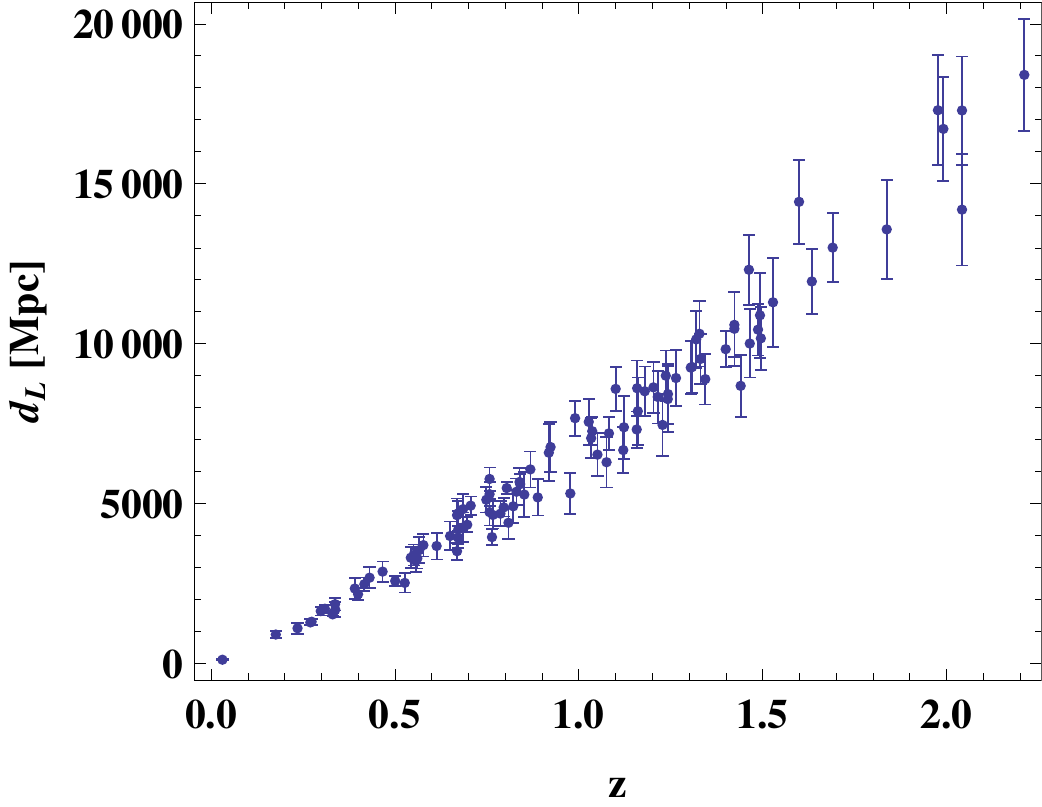}
	\includegraphics[width=0.315\textwidth, height=0.275\textwidth]{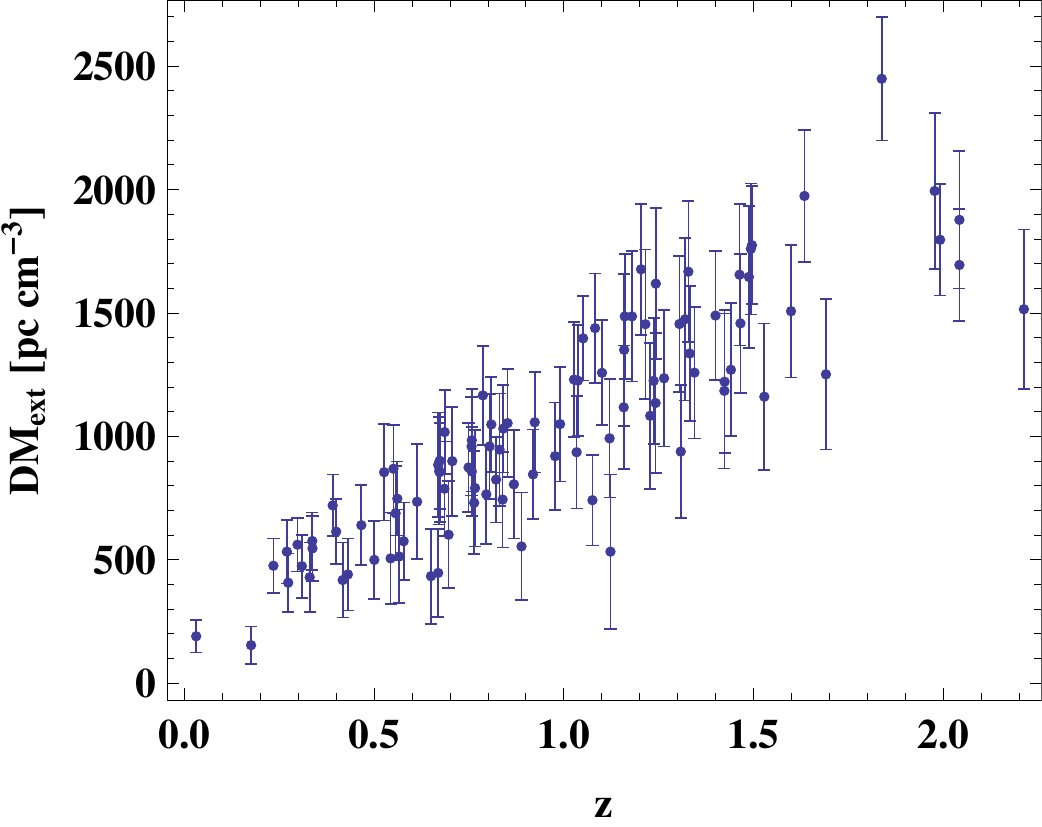}
	\includegraphics[width=0.33\textwidth, height=0.281\textwidth]{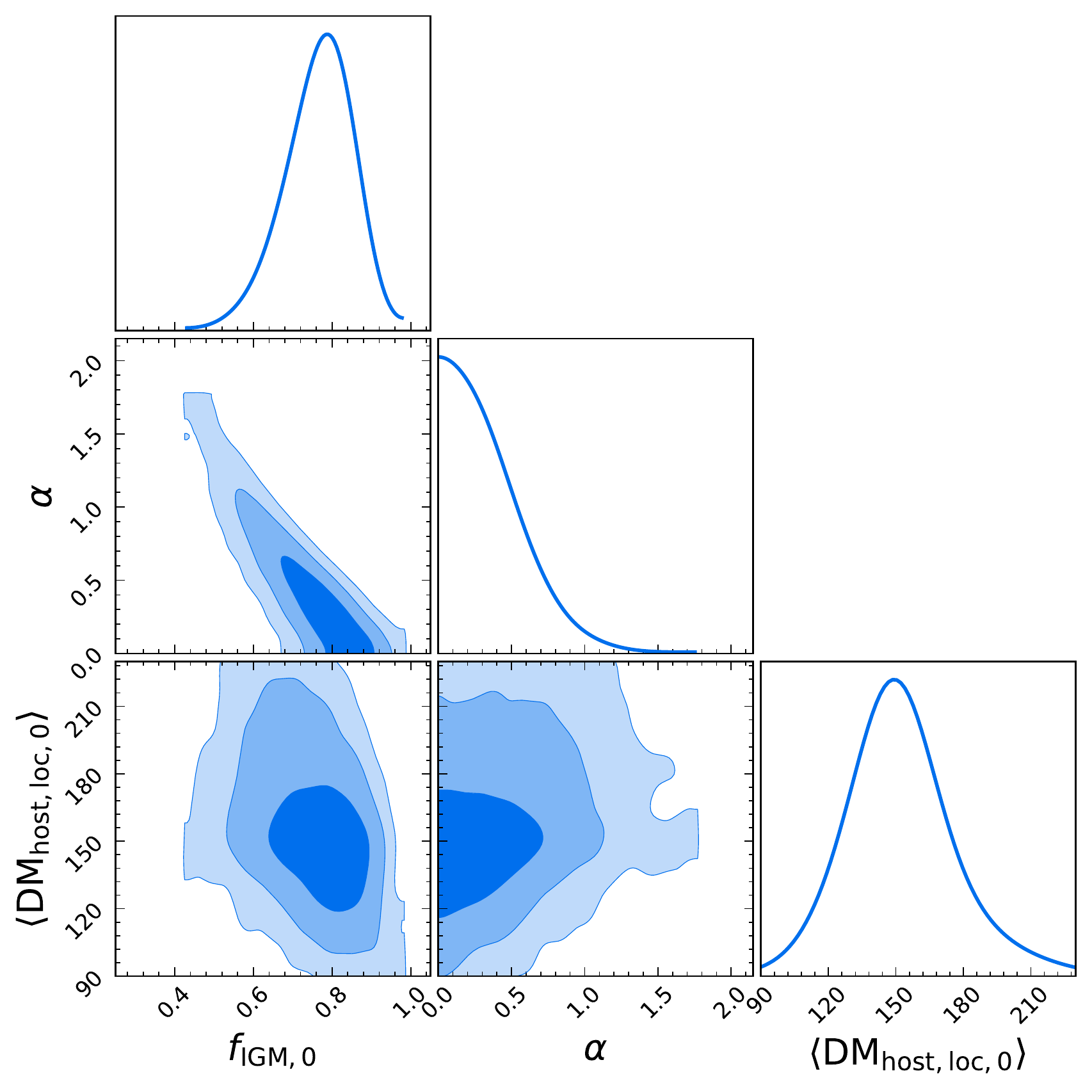}
	\caption{\label{Fig4} {\bf Left: } Mock data of 100 luminosity distances. {\bf Middle: } Mock data of 100 extragalactic dispersion measures. {\bf Right:} Constraints on $f_{\mathrm{IGM, 0}}$, $\alpha$, and $\langle\mathrm{DM}_{\mathrm{host, loc, 0}}\rangle$ from the corresponding simulations.}
\end{figure*}

\begin{figure*}[htbp]
	\centering
	\includegraphics[width=0.315\textwidth, height=0.275\textwidth]{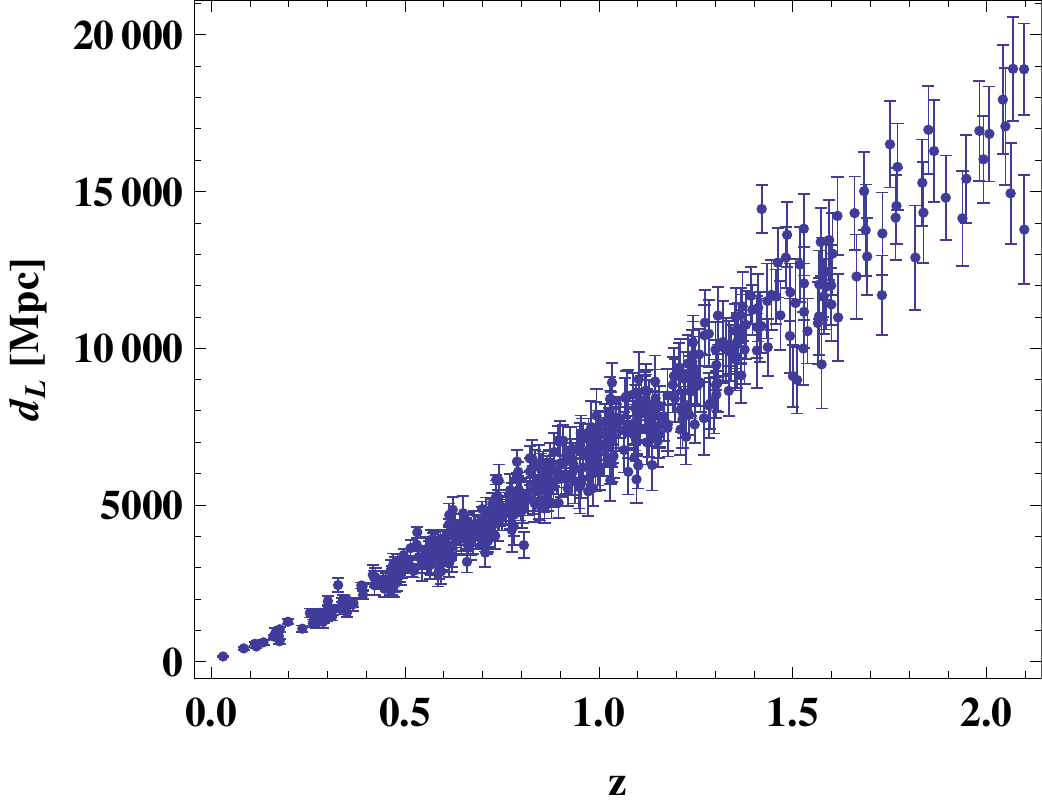}
	\includegraphics[width=0.315\textwidth, height=0.275\textwidth]{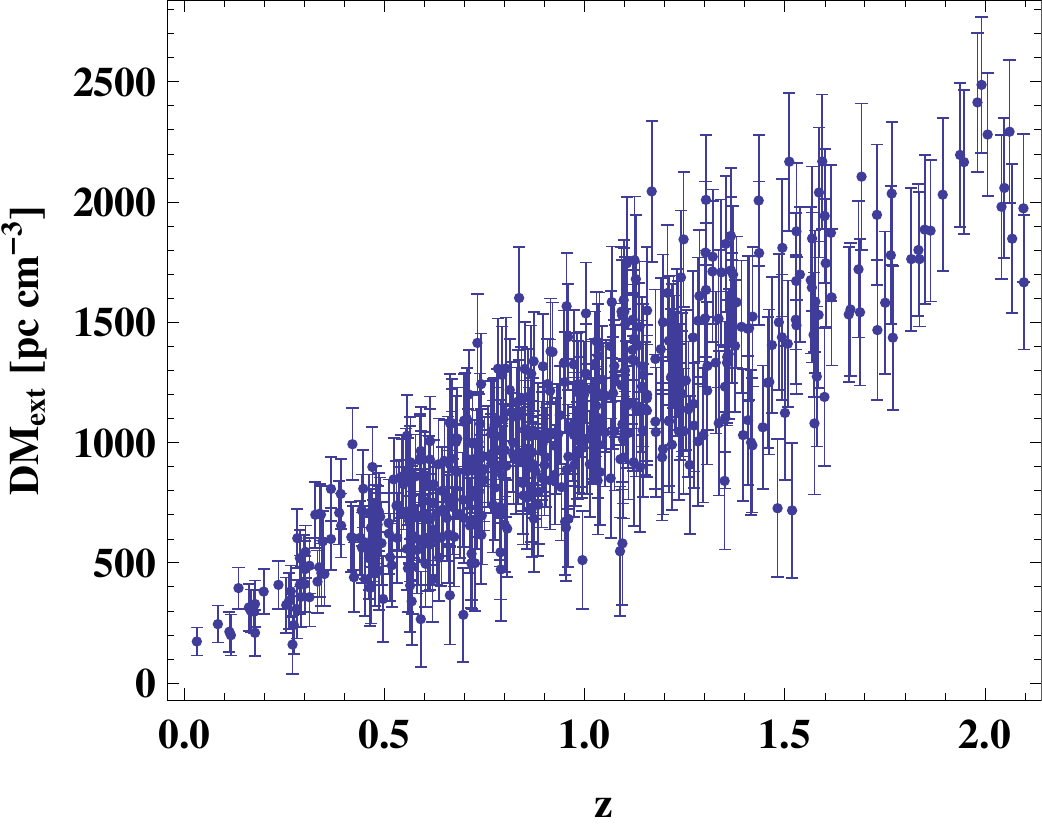}
	\includegraphics[width=0.33\textwidth, height=0.281\textwidth]{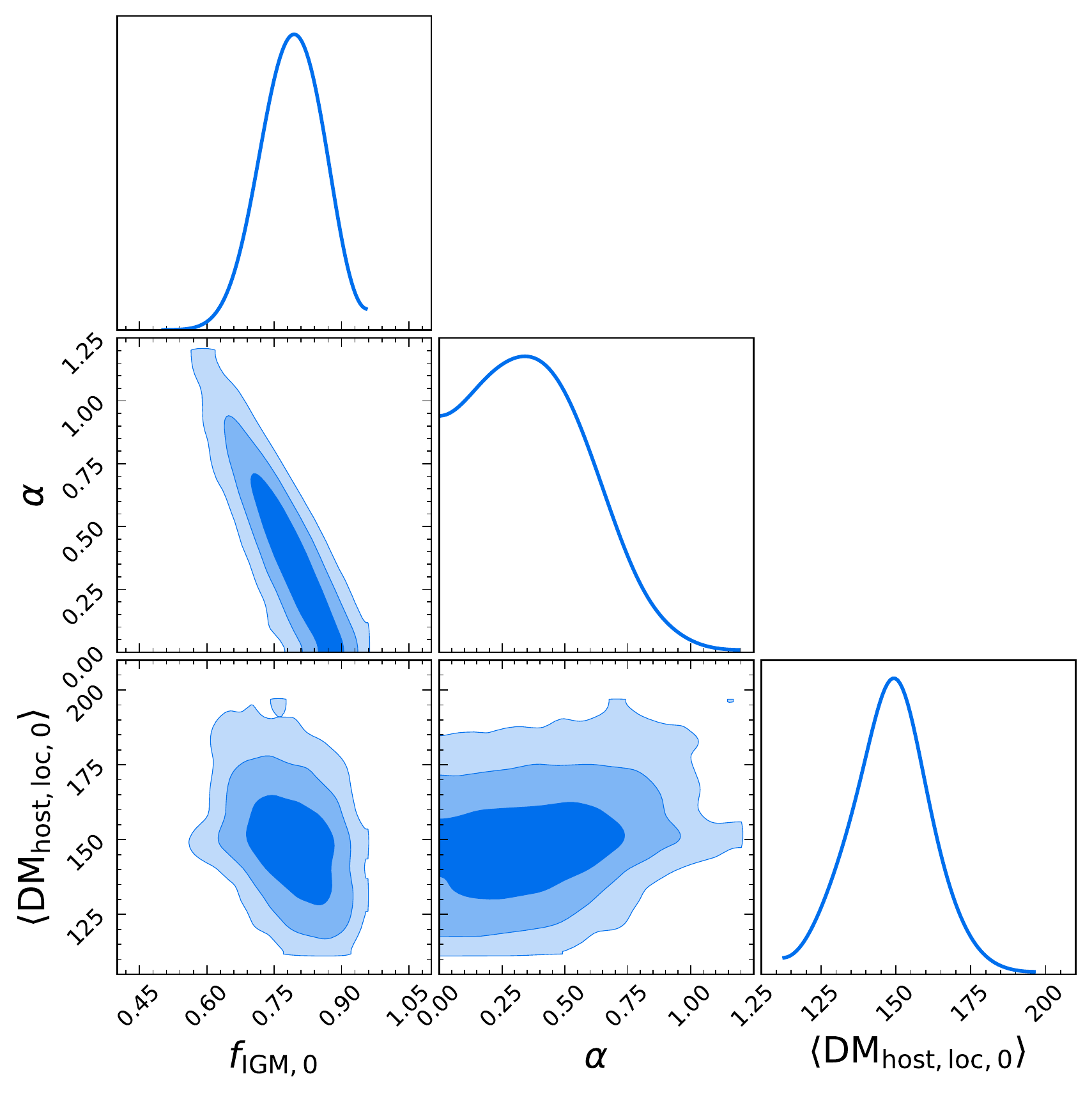}
	\caption{\label{Fig5} {\bf Left: } Mock data of 500 luminosity distances. {\bf Middle: } Mock data of 500 extragalactic dispersion measures. {\bf Right:} Constraints on $f_{\mathrm {IGM, 0}}$, $\alpha$, and $\langle\mathrm{DM}_{\mathrm{host, loc, 0}}\rangle$ from the corresponding simulations.}
\end{figure*}

Next, we carry out a similar analysis by considering a sample consisting of 100 joint measurements. Simulations for $d_{\rm{L}}$ and ${\rm {DM}}$ are shown in the left and middle panels of Figure \ref{Fig4}, respectively. Distributions of fitting results for both $f_{\rm {IGM, 0}}$, $\alpha$, and ${\rm \langle{DM_{host, loc, 0}}}\rangle$ from $\sim10000$ repeating simulations are presented in the right panel of Figure \ref{Fig4}. We obtain $f_{\rm {IGM, 0}}=0.81^{+0.07}_{-0.09}$, $\alpha=0.02^{+0.49}_{-0.02}$ and ${\rm \langle{DM_{host, loc, 0}}}\rangle=149.92^{+19.16}_{-15.40}$. Finally, we investigate the case with 500 joint measurements. Simulations for $d_{\rm{L}}$ and ${\rm {DM}}$ are shown in the left and middle panels of Figure \ref{Fig5}, respectively. Corresponding distributions of the fitting results from $\sim10000$ repeating simulations are presented in the right panel of Figure \ref{Fig5}. We obtain $f_{\rm {IGM, 0}}=0.81^{+0.06}_{-0.06}$, $\alpha=0.36^{+0.24}_{-0.23}$, and ${\rm \langle{DM_{host, loc, 0}}}\rangle=149.12^{+10.23}_{-12.77}$. These results suggest that, all concerning parameters in this method are unbiasedly inferred in a cosmology-independent manner when $\sim\mathcal{O}(10^2)$ such joint measurements are collected. 

\section{Conclusions and Discussions}\label{sec3} 
In this paper, we propose that with joint measurements of luminosity distance  and dispersion measure for FRBs, $d_{\rm {L}}/{\rm {DM}}$ can be a powerful probe for estimating the fraction of baryon mass in the intergalactic medium (IGM), and the local value of the host-galaxy DM. The quantity $d_{\rm {L}}/{\rm {DM}}$ is practically insensitive to cosmological parameters. This merit enables us to estimate $f_{\rm {IGM}}$ in a cosmology-independent manner. Through Monte Carlo simulations, we show that $f_{\mathrm{IGM}}$ can be unbiasedly inferred from only 50 such joint measurements. For a larger sample with 500 joint measurements, all concerning parameters including $f_{\mathrm{IGM, 0}}$, $\alpha$, and $\rm {DM_{host, loc, 0}}$ can be simultaneously constrained.

The key of performing the tests presented here is to measure the redshift of a large sample of FRBs. The first possibility relies on the repeatability of  FRBs.  The second possibility relies on their associations with other detectable signals such as GWs \citep{totani13,zhang16,wang16}, GRBs \citep{zhang14}, or any other bright counterparts \citep{yi14, spitler16, lyutikov16,zhang17}. In any case, once $z$ is measured for an FRB, $d_{\rm L}$ is likely obtained from known standardizable distance indicators (such as SNe Ia) around the similar redshifts. Our method can be then applied. The derived $f_{\rm IGM}$ and $\rm {DM_{host, loc}}$ would bring important information about the distribution of matter in the universe as well as the local environment of FRBs.

\section*{Acknowledgments}
This work was supported by the National Natural Science Foundation of China under Grants Nos. 11505008, 11722324, 11603003, 11373014, and 11633001, the Strategic Priority Research Program of the Chinese Academy of Sciences, Grant No. XDB23040100.

\end{document}